\documentclass[suppldata]{interact}   % to remove Article History; Compiled...

\usepackage{epstopdf}% To incorporate .eps illustrations using PDFLaTeX, etc.
\usepackage[caption=false]{subfig}% Support for small, `sub' figures and tables

\usepackage[numbers,sort&compress]{natbib}% Citation support using natbib.sty
\bibpunct[, ]{[}{]}{,}{n}{,}{,}% Citation support using natbib.sty
\usepackage{amsmath, amsfonts, amssymb}   
\usepackage{comment}
\usepackage{graphicx}

\usepackage{amsthm}                     %%% puts a qed symbol at end of proof environment
\newtheorem{theorem}{Theorem}
\newtheorem*{theorem*}{Theorem}

\newtheorem{fact}[theorem]{Fact}

\usepackage{color}             %%% for colored fonts

\usepackage{pdflscape}         %%% for changing a page to landscape orientation

\usepackage{algorithm}   %%% default is 'ruled'; 'boxed' is ok
\usepackage{algorithmic}

\newcommand{\bd}{\boldsymbol}

\newcommand{\defi}{\stackrel{\text{def}}{=}}	
\newcommand{\ttplus}{\text{\sc{e}}}

\newcommand{\qqq}{\textcolor{red}{ \underline{\textbf{???}} }}

\begin{document}

%\linenumbers

%%%%%%%%%%

%\begin{comment}
%\if0\blind
%{
   \title{ Modeling Material Stress Using Integrated Gaussian Markov Random Fields\thanks{
      This work was performed at Los Alamos National Laboratory and funded through the Laboratory Directed Research and Development program via projects 20170033DR and 20150594ER.} 
   }
   
   \author{ %%%%%
      \name{
      Peter W. Marcy\textsuperscript{a}\thanks{CONTACT Peter W. Marcy. Email: pmarcy@lanl.gov}, 
      Scott A. Vander Wiel\textsuperscript{a},
      Curtis B. Storlie\textsuperscript{b},
      Veronica Livescu\textsuperscript{c}, and
      Curt A. Bronkhorst\textsuperscript{d}
      }
      \affil{
      \textsuperscript{a}Statistical Sciences Group (CCS-6), Los Alamos National Laboratory (LANL), Los Alamos, NM, USA; 
      \textsuperscript{b}Mayo Clinic, Rochester, MN;
      \textsuperscript{c}Materials Science in Radiation and Dynamics Extremes (MST-8), LANL;
      \textsuperscript{d}Fluid Dynamics and Solid Mechanics Group (T-3), LANL.
      %\textsuperscript{d}Department of Engineering Physics, University of Wisconsin - Madison.
      } 
   } %%%%%
   
   %\date{2019}
   
   \maketitle

\bigskip

\begin{abstract}

The equations of a physical constitutive model for material stress within tantalum grains were solved numerically using a tetrahedrally meshed volume.  
The resulting output included a scalar vonMises stress for each of the more than 94,000 tetrahedra within the finite element discretization.
In this paper, we define an intricate statistical model for the spatial field of vonMises stress which uses the given grain geometry in a fundamental way.
Our model relates the three-dimensional field to integrals of latent stochastic processes defined on the vertices of the one- and two-dimensional grain boundaries.  
An intuitive neighborhood structure of said boundary nodes suggested the use of a latent Gaussian Markov random field (GMRF).
However, despite the potential for computational gains afforded by GMRFs, the integral nature of our model and the sheer number of data points pose substantial challenges for a full Bayesian analysis.
To overcome these problems and encourage efficient exploration of the posterior distribution, a number of techniques are now combined: parallel computing, sparse matrix methods, and a modification of a block update strategy within the sampling routine.
In addition, we use an auxiliary variables approach to accommodate the presence of outliers in the data. 

%Robust linear model \qqq The presence of outliers calls for a robust model, namely one with a Student-$t$ error distribution.

%We focus on the description and fitting of this unique model to the very large dataset but end by pointing out shortcoming related to unconditional realizations derived from it.

\end{abstract}

\noindent%
\vfill

%\newpage

%%%%%%%%%%

\section{Introduction}

Scientists who study the mechanics of materials at Los Alamos National Laboratory are interested in the fracture properties of real as well as simulated tantalum.
Physical experiments conducted at the macro-scale (i.e., much larger than single-crystal length) provide loading conditions to be used in meso-scale (where single crystals can be resolved) computational dynamic models of this material.
The need for the computational models stems from the following reasoning.
The outcome of the physical shock-loading experiments are tantalum plates with nucleated pores; the plates provide information on the end result, but provide no insight as to the mechanics of the process leading to the damage.
Because the initiation mechanisms are not directly measurable but still of interest, scientists have developed constitutive models of the internal stress-strain relationships to partly bridge this gap.
Owing to different phenomenology, models exist at both the meso- and macro-scales.
The ultimate goal of the researchers is to link meso- and macro-scale computational models in order to understand pore nucleation in tantalum.
A necessary first goal is to understand the spatial distribution of stress within the idealized meso-scale tantalum immediately preceding damage formation.

This paper represents the first step in a recent collaboration of theoretical materials scientists and statisticians to better understand the non-stationary spatial distribution of stress within a simulated collection of tantalum grains that have been subjected to shock-loading conditions. 
The materials scientists initially sought out statisticians due to the fact that many material properties and mechanisms are statistical in nature.
Further, while the materials modelers are able to write constitutive equations for the stress-strain relationships within tantalum, these equations do not naturally describe the spatial variability throughout a representative volume, and in particular, near grain boundaries.
Because the three-dimensional distribution of stress throughout the grain network is important for the understanding of damage initiation, the authors have begun to explore \emph{empirical}, data-driven (meta-)models.
This work represents our initial attempt to parameterize the tantalum stress fields through a nontraditional spatial statistical model.
Given the uniqueness of the dataset, the novel model it suggests, and the variety of practical computational techniques necessary to fit said model, we think this paper will be of general interest to a wider applied statistical audience.

While there are many potential flavors of spatial model to choose from, there are three important aspects of the simulated tantalum dataset that narrow the search.
First is the natural \emph{neighborhood structure} provided by the finite element techniques used to calculate the stress:  %(more on this in Section \ref{sec:MaterialScience}) 
two spatial locations can be considered neighbors if their volume %tetrahedral 
elements touch.
The second consideration is the \emph{sheer amount of data}:
the huge amount of output produced by the materials codes is an immediate and sizable hurdle for the fitting of any potential spatial statistical model.
The third consideration is the \emph{non-stationary} nature of the stress fields: 
variability tends to increase near grain boundaries.
All three of these aspects point to the use of Gaussian Markov random fields (GMRFs) for the tantalum stress fields.
GMRF models can use a neighborhood structure to specify the entries of the precision (as opposed to the covariance) matrix.
Not having to repeatedly invert the covariance matrix then leads to computationally efficient inference for large and possibly non-stationary spatial datasets \citep{Rue2005, Lindgren2011}.
%
%One type of model used in big spatial data problems is the Gaussian Markov random field (GMRF) due to its flexibility and computationally efficient inference \citep{Rue2005, Lindgren2011}.
%

On the practical side, it is possible to fit a wide class of (latent) GMRF models using the \verb1R-INLA1 software.
This \verb1R1 package facilitates approximate Bayesian inference using an integrated nested Laplace approximation \citep{Rue2009}.
However, this software cannot be used when there are more than four hyperparameters and/or the likelihood is sufficiently complicated.
In addition, the methodology of \cite{Lindgren2011} cannot be used for non-stationary models when lacking knowledge of stochastic partial differential equations (SPDEs).

In this paper we define a unique and intricate Bayesian non-stationary GMRF model for simulated tantalum which precludes the use of \verb1R-INLA1 and does not use SPDEs. 
It is based on an equation relating the 3D stress field to integrals of latent stochastic processes defined on the vertices of 2D and 1D grain boundaries, and its Bayesian inference requires modified sampling strategies within the Markov chain Monte Carlo (MCMC) routine.
The modifications we propose are directly generalizable to other high-dimensional GMRFs whose inference necessitates MCMC.

We start by providing more details of the materials science background in Section \ref{sec:MaterialScience}.
It will provide much needed context for the structure and nature of the dataset.
Next we specify our intricate model in Sections \ref{sec:Likelihood} and \ref{sec:Priors}.
The sampling routines for a full Bayesian analysis are presented in Section \ref{sec:Computation}, and results of the analysis are presented in Section \ref{sec:Results}.  
We end with conclusions and comment on the need for extensions.

%In this paper we...
%--- describe the unique/novel model for this unique dataset
%--- describe the fitting process
%--- gripe about making new realizations
%--- comment on reformulations/extensions

\section{Computational Materials Science Background}   \label{sec:MaterialScience}

The data we analyze come from a physical model (to be distinguished from our statistical ``meta"-model) of how tantalum crystallites (grains) respond to stress loading conditions.  
Specifically, it is a thermo-mechanically coupled elasto-viscoplastic single-crystal constitutive model in which the dominant physical feature is the interaction of crystals within the polycrystal volume.
A more in-depth description of the mathematical equations can be found in \cite[Sec. 3.2]{Bronkhorst2007}.
The collection of coupled differential equations were solved numerically on a representative volume of tantalum grains using Abaqus 6.12 \citep{Abaqus2012}.

\begin{figure}[h!t]   \begin{center}
\includegraphics[width=5.0in]{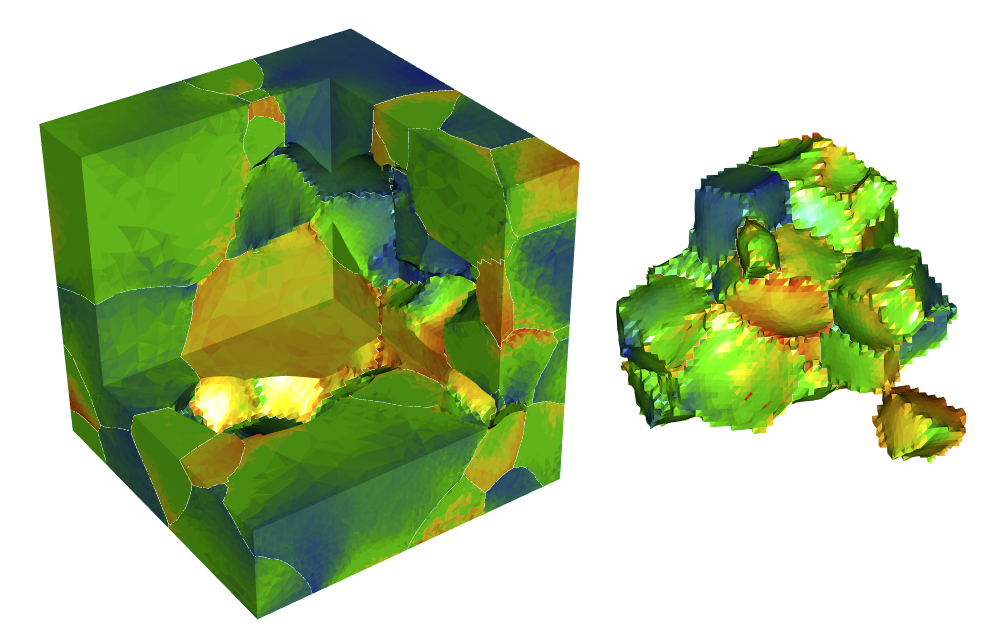}
%\includegraphics[height=3.2in]{70_grain-cutout.pdf}
%\hspace*{-1.7in}
%\includegraphics[height=2.9in]{18_grain.pdf}
\caption{A cutout of the full representative volume (left) showing the subset of 18 fully internal grains (right).  Blue indicates low ($>$535 MPa) and red indicates high ($<$1147 MPa) vonMises stress.}
\label{fig:70to18Grains}
\end{center}   \end{figure}

%\begin{figure}[h!t]   \begin{center}
%\includegraphics[height=3.0in]{18_grain-cutout.pdf}
%\caption{\redud{placeholder for ???}}
%%\label{fig:18GrainTriangulation}
%\end{center}   \end{figure}

\begin{figure}[h!t]   \begin{center}
\includegraphics[width=3.0in]{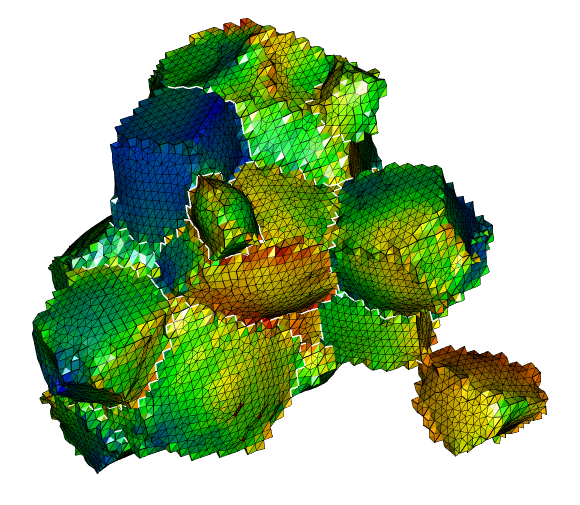}
\caption{The 18 grain volume with boundary triangulations visible.}
\label{fig:18GrainTriangulation}
\end{center}   \end{figure}

\begin{figure}[h!t]   \begin{center}
\hspace*{-32pt}  \includegraphics[width=6.5in]{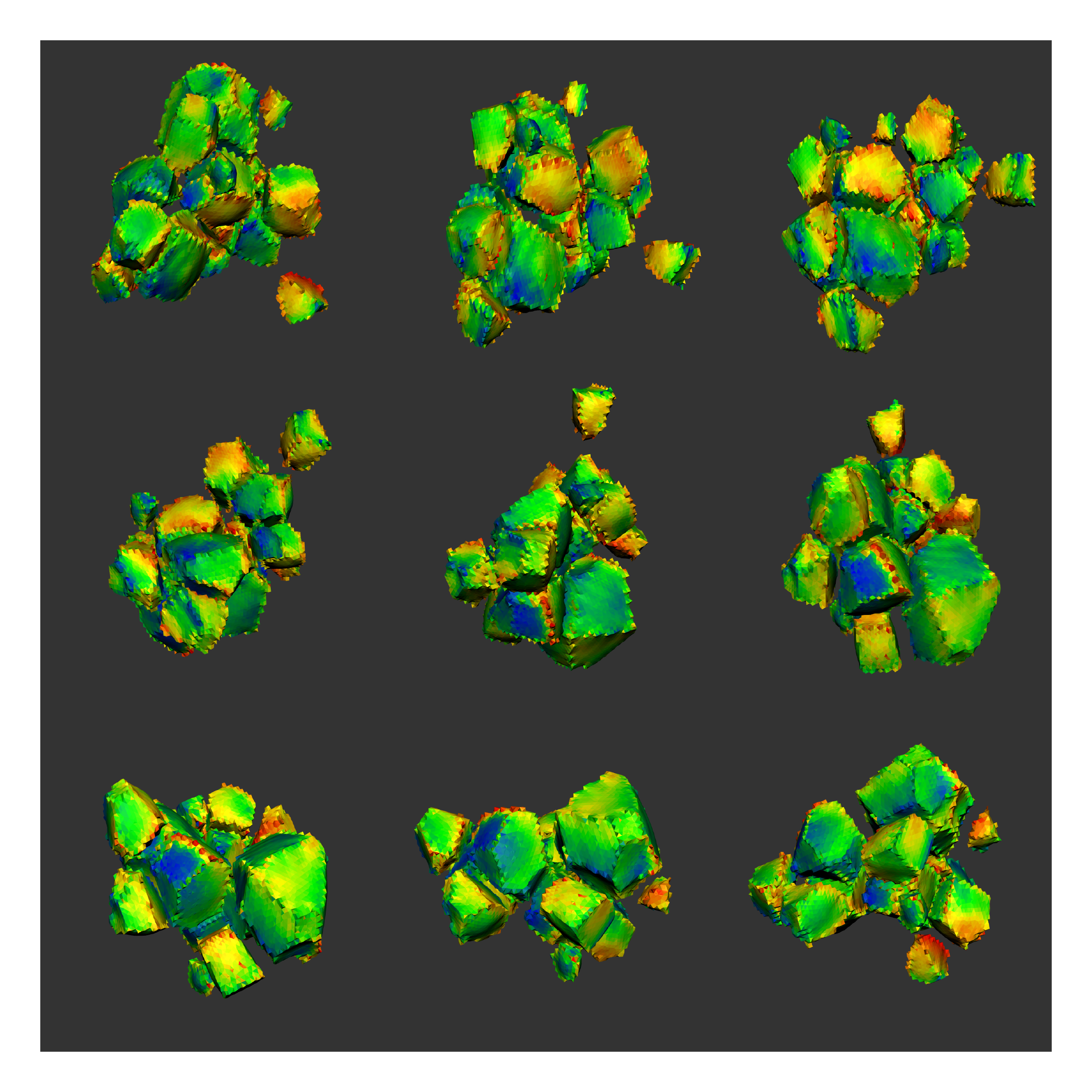} 
\caption{ The 18 grains partially separated and rotated about the spatial $x=y$ axis to show variability of vonMises stress on grain boundaries; 
the progression goes from the top-left to the bottom-right panel.  
(The spatial axes are defined by the cube faces of Figure \ref{fig:70to18Grains} with the $z$ axis being the vertical.) }
\label{fig:Movie}
\end{center}   \end{figure}

The finite element discretization and boundary conditions used by the numerical solvers were formulated using experimental measurements.
A representative three-dimensional polycrystalline cube (sidelength 165 $\mu$m) with 70 tantalum grains was generated with an open-source software utility Dream3D \citep{Groeber2014} using information derived from two-dimensional electron-backscatter diffraction data of \cite{Bronkhorst2007}.  
In general, the experimental orientation image maps from the three primary orthogonal directions of a material are used by Dream3D to train a statistical model of the grain size, shape, and orientation for the microstructure.  
This model can be used by Dream3D to generate as many statistically equivalent microstructures as needed.  
In our case, the resulting volume contained 577,445 conformally meshed tetrahedral computational elements on 106,904 vertices/nodes. 
The finite elements along grain boundaries are smaller in order to provide a higher resolution of the spatio-temporal mechanics in these regions.
The meshing algorithms within Dream3D are not perfect and are constantly being refined by the developers.
For instance, some jagged artifacts of the conformal meshing can be seen where three grains meet (Figures \ref{fig:70to18Grains}, \ref{fig:18GrainTriangulation}, and \ref{fig:Movie}), but this was the state of our spatial data at the time of generation.

The time-dependent boundary conditions for the compressive forces acting perpendicular to the cube-faces of the computational polycrystal were derived from macro-scale simulations of plate impact experiments of \cite{Bronkhorst2016}.  
The loading history was imposed up to the point at which the simulations suggested pore nucleation in the tantalum-on-tantalum plate experiments.  
%\redud{Specific loading conditions??}
This is the sole time point for which we consider the output.

The output of Abaqus includes many state variables including the tensor-valued quantities: strain, strain rate, and stress.
In this study, we consider only vonMises stress (in units megapascals, MPa), a scalar reduction of the Cauchy stress tensor, because of its direct relevance to material yield and damage initiation \citep{Ottosen2005}.
%From the Cauchy stress tensor $\bd\sigma$
%$$  2 \sigma^2_{vM} = (\sigma_{11} - \sigma_{22})^2 + (\sigma_{11} - \sigma_{33})^2 + (\sigma_{22} - \sigma_{33})^2   +   6(\sigma^2_{12} + \sigma^2_{13} + \sigma^2_{23})  $$

\begin{figure}[h!t]   \begin{center}
\includegraphics[width=4.0in]{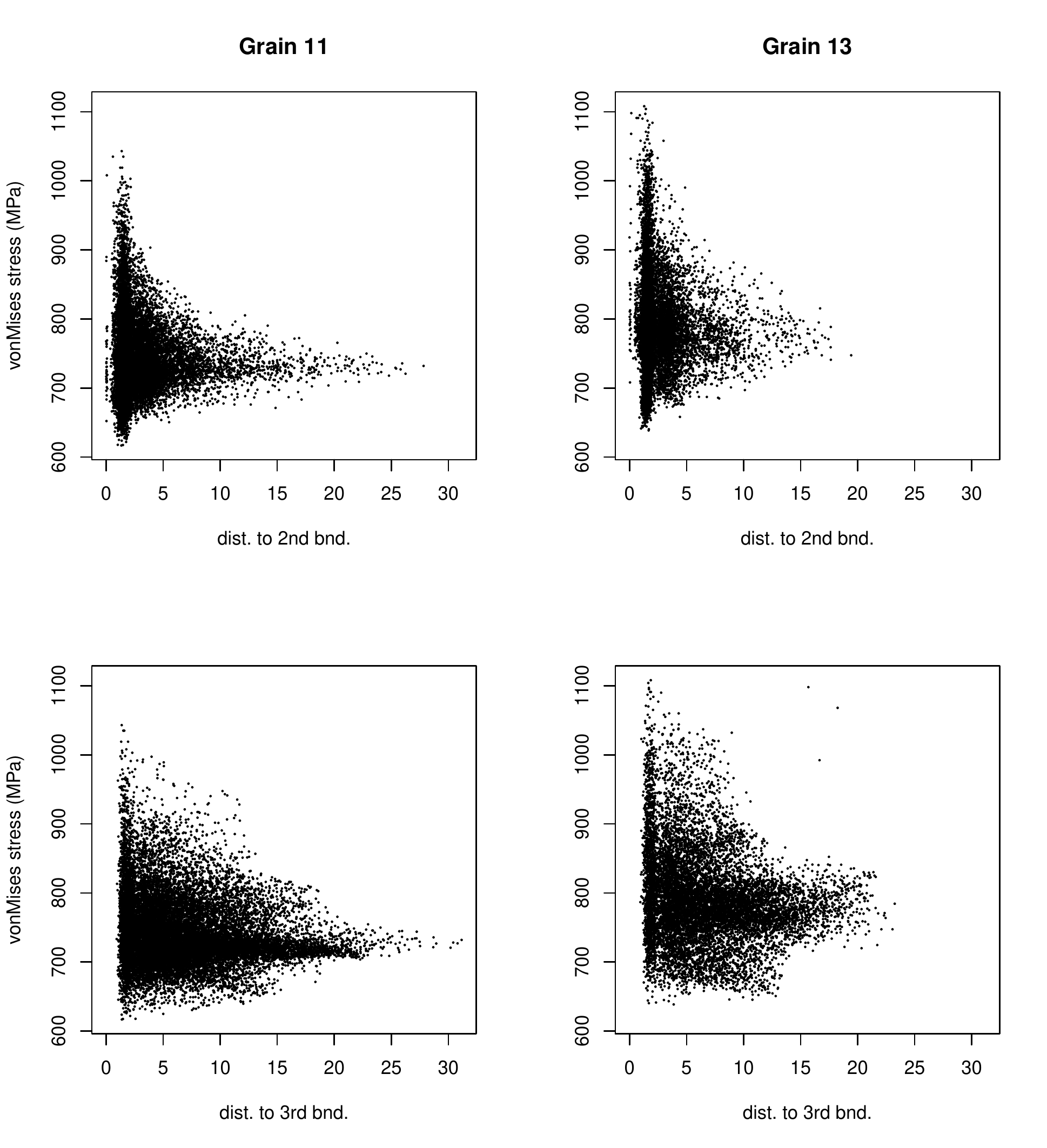}
\caption{ The variability of vonMises stress decreases as distance from boundaries increases. }
\label{fig:vonMisesDecay}
\end{center}   \end{figure}

%The spatial domain was subsetted to reduce the sample size as well as remove potential boundary effects induced by the forces acting upon the cube's faces.
The spatial domain was subsetted to remove potential boundary effects induced by the forces acting upon the cube's faces.   %%% Storlie
We consider the 18 complete grains which do not intersect any cube-face; these internal grains contain a still formidable 94,274 tetrahedral elements on 60,966 nodes (10,851 of these being boundary nodes)
%The subsetted data are shown in Figure \ref{fig:70to18Grains}.
The subsetted data are shown in Figures \ref{fig:70to18Grains} and \ref{fig:Movie}.

Initial explorations of the stress field revealed a few important features to include in our statistical model.  
First, grains tend to have their own baseline vonMises stress, suggesting a grain-specific mean.
Next, there is clear spatial correlation with grain boundaries appearing to generate areas of both high and low stress.
However, the effect of grain boundaries is not entirely consistent in that some boundaries can be seen to separate areas of relatively similar stress while others separate high and low stress regions.
A decay in variability as a function of distance to second- and third-order grain boundaries can be seen in Figure \ref{fig:vonMisesDecay}.
(Second-order boundaries are surfaces between two grains while third-order boundaries are intersection curves between three grains.)
Finally, there are outliers of (usually high) stress which are unlike surrounding values -- this suggests that any reasonable statistical model will need an error distribution with heavy tails.  
% hypothesize reason \qqq
In the next section we detail a model that can accommodate all the above features.

\section{Bayesian Hierarchical Model: Likelihood}   \label{sec:Likelihood}

We now define a scalar random field model for vonMises stress throughout the 18-grain volume.  
Let $Y(\bd s)$ be the vonMises stress at spatial location $\bd s$, and let $g \equiv g(\bd s)$ be the grain at location $\bd s$.  
The idealized version of our model relates the expected stress at a location $\bd s$ in three-dimensional space to averages of unobserved functions of one and two dimensions: 
\begin{align}
   E \left\{ Y(\bd s) \right\}    &=   \mu_g \ + \ \int_{\widetilde B_g}  e^{-\phi_\beta d(\bd s , \bd v)}  \beta_g (\bd v) \ d \bd v
                       \quad + \ \int_{\widetilde C_g}  e^{-\phi_\gamma d(\bd s , \bd v)}  \gamma_g (\bd v) \ d \bd v   \label{eq:ModelIdeal}
\end{align}
where $\mu_g$ is the baseline stress in grain $g\in\{1,\ldots G\}$, $\phi_\beta, \phi_\gamma > 0$, and $d(\bd s , \bd v)$ is the Euclidian distance between spatial locations $\bd s$ and $\bd v$ in $\mathbb{R}^3$.  
$\widetilde B_g$ is the second-order boundary of grain $g$, i.e., the two-dimensional manifold where grain $g$ contacts neighboring grains, and is the domain of the unobserved function $\beta_g(\cdot)$.  
Hence, the integral over $\widetilde B_g$ is a surface integral.  
Similarly, $\widetilde C_g$ represents the third-order boundary of grain $g$, i.e., the union of one-dimensional manifolds where grain $g$ simultaneously contacts any two (or more) neighboring grains.  
It is the domain of the function $\gamma_g(\cdot)$, and the corresponding integral is therefore a sum of line integrals.

\begin{figure}[h!t]   \begin{center}
\includegraphics[width=5.0in]{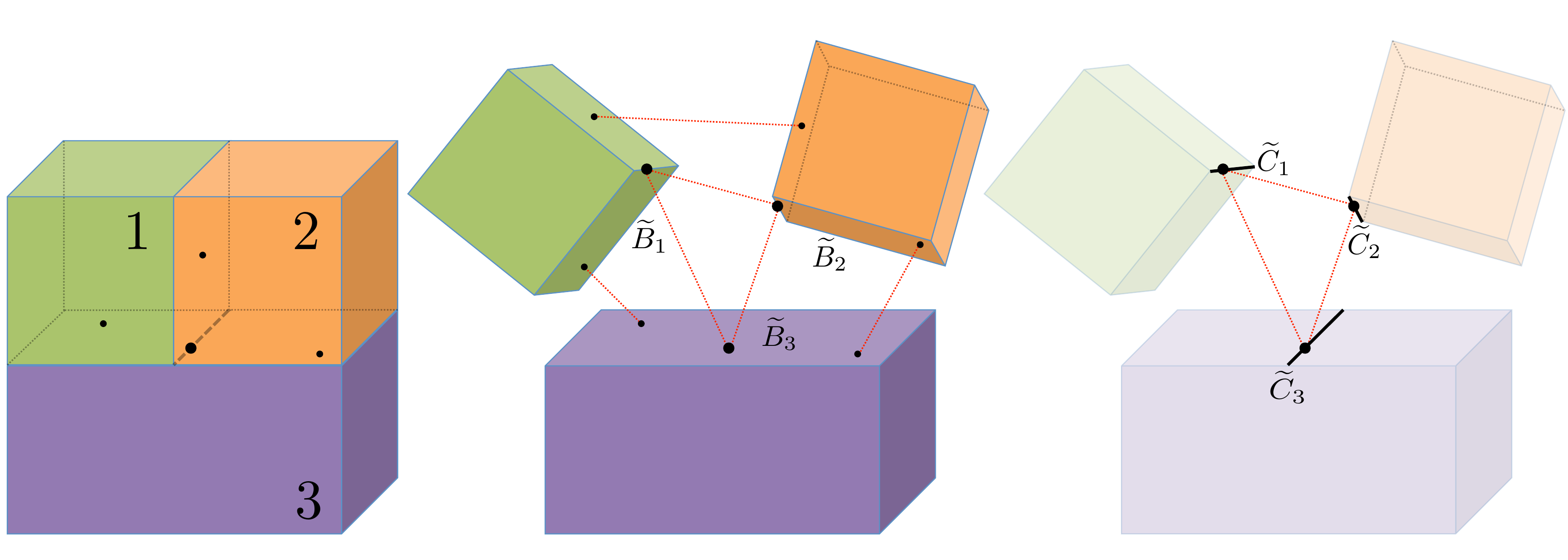}
\caption{ A simple geometry of three grains (left panel).
The second- and third-order boundary domains for the integrals of (\ref{eq:ModelIdeal}) are displayed in the center and right panels (respectively).  
The red lines connect the multiple representations of the original four boundary points. }
\label{fig:CartoonGrains}
\end{center}   \end{figure}

Figure \ref{fig:CartoonGrains} displays domains $\widetilde B_g$ and $\widetilde C_g$ for a simple geometry of three grains.  
Each point on a second-order boundary which is not a higher-order boundary (e.g., the smaller dots in the left panel of the figure) is represented exactly twice in the integrals of (\ref{eq:ModelIdeal}); 
each point on a third- or higher-order boundary (e.g., the larger dot) is represented six times.
The red lines in the center and right panels of the figure connect the multiple representations.
Having distinct functions defined on a common boundary allows for different local spatial behavior on either side of the boundary.
Furthermore, the exponential kernel in the integrals of (\ref{eq:ModelIdeal}) permits a distance weighted averaging that can capture decaying trends such as those in Figure \ref{fig:vonMisesDecay}.

%\begin{figure}[h!t]   \begin{center}
%\includegraphics[width=2.0in]{Grain_Boundary_2.pdf}
%\includegraphics[width=2.0in]{Grain_Boundary_3.pdf}
%\caption{\redud{placeholder for ???}}
%\label{fig:CartoonGrainBoundaries}
%\end{center}   \end{figure}

The relation defined in (\ref{eq:ModelIdeal}) is clearly an integral equation, and for this reason is ``idealized" because its parameter set is actually a collection of functions $\{ \beta_g(\bd v) , \gamma_g(\bd v) \}_{g=1}^G$.  
In fact, from a mathematical perspective, it is a Fredholm integral equation of the first kind \citep[Ch. 8]{Wahba1990}, and because the kernel is symmetric, this problem of recovering the unknown functions is called \emph{deconvolution}.  
Furthermore, when the kernel parameters $\phi_\beta$ and $\phi_\gamma$ are not given, the scenario is referred to as blind deconvolution \citep{Wipf2014}.
From a statistical perspective, by assuming the functions are stochastic processes (and being careful with the definition of the integral) one specifies a ``kernel mixing" or ``process convolution" model \citep{Higdon2002, Banerjee2015}.  
Our model in (\ref{eq:ModelIdeal}) differs from the typical process convolution because the domain of integration is of lower dimension than the spatial domain; 
the expected stress throughout the volume will be determined completely by the behavior at the one- and two-dimensional grain boundaries.

Our actual model utilizes the discrete geometry from the finite element approximation.  
The spatial location of a particular observed datum is taken to be the centroid of its tetrahedral computational element, i.e. the coordinate average of the four vertices/nodes.  
Moreover, the entire representative volume (cube of 70 grains) is \emph{conformally} meshed with tetrahedral elements so that the region where two grains meet is a common collection of triangles.  
That is, each two-dimensional manifold $\widetilde B_g$ is approximated by a set of nodes $B_g$ belonging to tetrahedral faces of the boundary.  
The region where three grains meet is a line segment defined by a collection of common vertices, and all such line segments for a grain form $C_g \approx \widetilde C_g$.  
Note that $C_g \subset B_g$.
Consequently, the process $\beta_g$ (and $\gamma_g$) will be characterized at the discrete node points in $B_g$ ($C_g$) to carry out a quadrature approximation of (\ref{eq:ModelIdeal}).

The model for vonMises stress of element $m \in \{1, \ldots , M=94,274\}$ having centroid $c_m$ within grain $g \in \{1,\ldots, G=18\}$ is given by 
%\footnotesize
\begin{align}    
   Y_m   &=   \mu_g + 
   \sum_{v \in B_g} \beta_g (v) e^{-\phi_\beta d(c_m , v)} \Delta v   +  
   \sum_{v' \in C_g} \gamma_g (v') e^{-\phi_\gamma d(c_m , v')} \Delta v'   +
   \varepsilon_m   
\\
%\end{align}
%with differentials
%\begin{align}
  \Delta v  &= (1/3) \cdot \text{total \emph{area} of all second-order tetrahedral faces on $v$}  \\
  \Delta v' &= (1/2) \cdot \text{total \emph{length} of all third-order tetrahedral edges on $v'$}  \nonumber   \\
\varepsilon_m   &\stackrel{iid}{\sim}   t_{df}(0, \sigma^2)   \nonumber
\end{align}
%\normalsize
%$\varepsilon_{\bd s} \stackrel{iid}{\sim} t_{df}(0, \sigma^2)$ represents independent additive (heavy-tailed) noise.  
%
%The integrals in (\ref{eq:ModelIdeal}) can be computed with a quadrature approximation using the discretizations
%\begin{align}    
%   \sum_{\bd u \in B_g} \beta_g (\bd u) e^{-\phi_\beta d(\bd s , \bd u)} \Delta\bd u   \quad &\text{ and } \quad
%   \sum_{\bd v \in C_g} \gamma_g (\bd v) e^{-\phi_\gamma d(\bd s , \bd v)} \Delta\bd v   \ 
%\end{align}
%with
Let $\bd\beta_g$ represent a column vector of the elements in $\{\beta_g(v) : v \in B_g\}$, sorted by increasing node index.  
Similarly, let $\bd\gamma_g$ represent the sorted vector of elements in $\{\gamma_g(v) : v \in C_g\}$.
Let $\bd\beta = (\bd\beta_1^\top, \ldots, \bd\beta_G^\top)^\top$ and $\bd\gamma = (\bd\gamma_1^\top, \ldots, \bd\gamma_G^\top)^\top$ be the combined vectors of second- and third-order grain boundary processes for all grains; dim$(\bd\beta)=14,434$ and dim$(\bd\gamma)=3,482$.  
This defines a vectorized form of the model:
\begin{align} 
   \bd y   \ &= \   \bd\mu \ + \ \bd X_b \bd\beta \ + \ \bd X_c \bd\gamma \ + \ \bd\varepsilon   \\
   \bd\mu   &=   (\mu_1 \bd 1_{M_1}^\top , \ldots , \mu_G \bd 1_{M_G}^\top)^\top   \nonumber   \\
   [\bd X_b]_{m,p}   &= 
   e^{ -\phi_\beta d( c_m, v_{n(p)} ) }  \Delta v_{n(p)} \ \cdot \ \mathbb{I} \{ v_{n(p)} \in B_{g(m)} \}   \nonumber
\end{align}
where $\bd 1_k$ is a $k$-vector of 1's; $\sum_{g=1}^G M_g = M$;  $g(m)$ gives the grain index of element $m$ and $n(p)$ gives the second-order boundary node index corresponding to $\bd\beta$-index $p$; $\mathbb I \{\cdot \}$ is an indicator function.  
The matrix $\bd X_c$ is defined similarly; note that the indicator function makes $\bd X_b$ and $\bd X_c$ block diagonal.  
By definition, a node in a third-order boundary is also a node in a second-order boundary, implying that columns of $\bd X_c$ will be columns of $\bd X_b$ for $\phi_\beta = \phi_\gamma$.
Multicollinearity is also expected within the columns of $\bd X_b$ because of the large number of small tetrahedra on the boundaries.
In the applied mathematics, the inverse problem is then said to be \emph{ill-conditioned} (on top of being ill-posed).
Hence, we will allow for correlated $\bd\beta$ entries through their prior.
More on this in Section \ref{sec:PriorBeta}.

Before we explicitly write the likelihood we need one more result.

\subsection{Auxiliary Variables for Heavy-Tailed Errors}

The Student-$t$ distribution can be interpreted as a scale-mixture of normal distributions due to the following result.
\begin{fact}
If $Z \sim N(0, \sigma^2)$  and  
$\omega \sim InvGam(\frac{df}{2}, \frac{df}{2})$, 
%\\ \hspace*{0.65in}
then the quantity
$\varepsilon \defi \sqrt{\omega} \cdot Z  \sim  t_{df}(0, \sigma^2)$
\end{fact}
\noindent
This implies that 
$\omega_m  \stackrel{iid}{\sim}  InvGam(\frac{df}{2}, \frac{df}{2})$ 
and 
$(\varepsilon_m \ | \ \omega = \omega_n)  \stackrel{iid}{\sim}  N(0, \omega_n \cdot \sigma^2) $ 
provide marginal Student-$t$ distributions: 
$\varepsilon_m   \stackrel{iid}{\sim}   t_{df}(0, \sigma^2)$.
Hence we obtain a regression model with heavy-tailed errors by introducing an auxiliary/latent-data vector $\bd\omega$ (having the same length as the original data $\bd y$) which can easily be updated with a Gibbs step \citep[e.g.,][Sections 12.1 and 17.2]{Geweke1993, Gelman2014}.
%Zellner 1976 ??? 

Finally, the likelihood for the parameters of our model is
\begin{align}
   L(\bd\mu , \phi_\beta , \bd\beta , \phi_\gamma , \bd\gamma , \sigma^2 , \bd\omega \ ; \ \bd y )   &\propto   |\bd W|^{1/2} \exp \left\{ -\frac{1}{2} (\bd y - \widetilde{\bd\mu})^\top  \bd W  (\bd y - \widetilde{\bd\mu}) \right\}   \label{eq:Likelihood} \\
   \widetilde{\bd\mu}   &\defi   \bd\mu + \bd X_b \bd\beta + \bd X_c \bd\gamma   \nonumber \\
   \bd W   &\defi   \frac{1}{\sigma^2} \text{diag}\left( \frac{1}{\omega_1},\ldots, \frac{1}{\omega_M} \right)   \nonumber
\end{align}
The matrices $\bd X_b$ and $\bd X_c$ depend on $\phi_\beta$ and $\phi_\gamma$, respectively.  
The priors for the parameters above will be given next.

\section{Bayesian Hierarchical Model: Priors}   \label{sec:Priors}

\subsection{Priors on $\beta$ and $\gamma$}   \label{sec:PriorBeta}

Most kernel convolution models feature a Gaussian mixing distribution, but extensions are possible (see e.g. the overview and references in \cite{Banerjee2015}).  
In particular, instead of assuming the elements of $\bd\beta$ are $iid$ normal, we will allow for spatial correlation (to counter the ill-conditioned nature of the convolution) by assuming the entire vector follows a multivariate normal distribution corresponding to a Gaussian Markov Random Field (GMRF).  
%identifiability of ill-conditionedness ???
%We do this because \qqq. More below.
Recent examples of GMRFs as priors in deconvolution-style inverse problems include \cite{Higdon2002, Lee2002, Lee2005, Bardsley2013, Zhang2013, Zhang2016}.
The random field models for $\bd\beta$ and $\bd\gamma$ will determine a field for $\bd y$ conditional upon $\bd X_b$ and $\bd X_c$.
In fact, the latent fields can be quite rough, and the resulting ``integrated" field for vonMises stress will be continuous, perhaps even differentiable.

A GMRF facilitates computational efficiency for large spatial models by being defined directly through a sparse precision (inverse covariance) matrix \citep{Rue2005, Lindgren2011}.  
GMRF models are naturally applicable when the data possess a neighborhood (vertex and edge) structure: the $ij$th precision entry will be non-zero if and only if the neighborhood includes edge $ij$.
In essence, the computationally efficient inference comes at the cost of having to be defined through \emph{conditional} moments.  This stands in contrast to traditional Gaussian process models which are easily specified through unconditional means and covariances, but whose associated parameter estimation is computational infeasible for large datasets.

The geometry of the tetrahedrally meshed 18-grain volume lends itself naturally to a GMRF on the second- and third-order boundaries.  
We describe the model for $\bd\beta$, and the extension to $\bd\gamma$ will be abundantly clear.
There can be many thousand second-order nodes in each $B_g$ set, and this means a large $\bd\beta_g \subset \bd\beta$.  
However, there is a very intuitive neighborhood structure: two nodes are neighbors iff they are vertices of a common tetrahedral element.  
For this reason $\bd\beta_g$ is assumed to be a GMRF on $B_g$, and further, that the entire collection of grain boundary processes $\{ \bd\beta_g \}_{g=1}^G$ are statistically dependent on one another through partial correlations at their shared boundaries.  
This is to say, the entire $\bd\beta$ vector is a GMRF whose definition necessitates further notation.
Let $B_{g,v} \subset B_g$ denote the neighbors of node $v$ on the boundary of grain $g$. 
Let $G_g$ denote the neighboring grains of grain $g$, and $G_{g,v} \subset G_g$ denote only those neighboring grains that also share node $v$, i.e., $G_{g,v} = \{g' : g' \in G_g  \text{ and } v \in B_{g'} \}$.

It is assumed that $\beta_g(v)$, conditional on all other elements of $\bd\beta$, only depends on 
$\{\beta_g(v') : v' \in B_{g,v}\}$ and 
$\{\beta_{g'}(v) : g' \in G_{g,v}\}$.
The first set represents within-grain neighbors (``wgn"): the same boundary process at physically adjacent nodes; 
the second set is the between-grain neighbors (``bgn"): different processes at the same node.
Again, refer to Figure \ref{fig:CartoonGrains} in which the bgn's are connected with red lines.
%Two entries $\beta_p$ and $\beta_q$ of $\bd\beta$ are within-grain neighbors iff  $p \neq q$, but $g_p=g_q$ and $v_p \in B_{g_p,v_q}$

The model for the $\beta$ random field ($\gamma$ and its parameters defined analogously) is 
\begin{align}
   \bd\beta   & \sim   N \left( \nu_\beta \bd 1, \ \bd Q_\beta^{-1}\right)   \\
   [\bd Q_\beta]_{p,q}   &= \theta_\beta \cdot
   \left\{ \begin{array}{ll}
      -1  &  \mbox{if wgn: }  p \neq q, \  g(p) = g(q), \  v_{n(q)} \in B_{g(p), v_{n(p)}}  \\
      -\rho_\beta  &  \mbox{if bgn: }  p \neq q, \  n(p) = n(q), \  g(q) \in G_{g(p), v_{n(p)}}  \\
      \frac{K_p}{\kappa_\beta}  &  \mbox{if } p = q  \\
      0 & \mbox{otherwise} 
   \end{array} \right.   \label{eq:PriorPrecision}   \\
   K_p   &=   \left( |B_{g(p),v_{n(p)}}|  + \rho_\beta  |G_{g(p), v_{n(p)}}| \right)   \nonumber
\end{align}
where $\nu_\beta$ is the mean of the $\beta$ process, constant over all grain boundaries.  
The cardinalities $|B_{g(p),v_{n(p)}}|$ and $|G_{g(p), v_{n(p)}}|$ are precisely the number of wgns and bgns of $\beta_p$, and so it can be observed that the matrix is diagonal-dominant when 
$\kappa_\beta \in (0,1)$ and 
\begin{align}
   \rho_\beta \in \left( -\underset{p}{\text{min}} |B_{g(p),v_{n(p)}}| / |G_{g(p), v_{n(p)}}| \ , \ 1 \right) \ .   \label{eq:RhoBounds}
\end{align}
Diagonal-dominance is a sufficient condition for the positive definiteness of $\bd Q_\beta$.  
A similar GMRF parameterization was advocated by \cite{Storlie2016} for use with multiple, distinct neighbor types.

%\redud{change the set B to be node indices not nodes ???}

Properties of GMRFs (e.g., Rue and Held (2005), Theorem 2.3) can be used to gain insight into the parameterization above:
\begin{align}
   E(\beta_p | \bd\beta_{-p})   &=   \nu_\beta + \frac{\kappa_\beta}{K_p} 
   \left( \sum_{v' \in B_{g(p) , v_{n(p)}}} \left( \beta_{g(p)}(v') - \nu_\beta \right)   \right.   \label{eq:CondMean}   \\ 
   & \hspace*{0.58in}   \left.   + \rho_\beta \sum_{ g' \in G_{g(p), v_{n(p)}}} \left( \beta_{g'}(v_{n(p)}) - \nu_\beta \right) \right)   \nonumber   \\
   Var(\beta_p | \bd\beta_{-p})   &=   \frac{\kappa_\beta}{\theta_\beta K_p}   \nonumber   \\
   Corr(\beta_p, \beta_q | \bd\beta_{-pq})   &=      
      \left\{ \begin{array}{rl}
         \kappa_\beta / \sqrt{K_p K_q}  &  \mbox{if wgn} \\
         \rho_\beta \kappa_\beta / \sqrt{K_p K_q}  &  \mbox{if bgn}
      \end{array} \right.   \nonumber
\end{align}
The parenthesized factor in (\ref{eq:CondMean}) is a weighted average of the neighbors, where $\rho_\beta$ controls the relative weight between the wgns and bgns.  
The $\kappa_\beta \in (0,1)$ then sets the proportion of the weighted average to use for the conditional auto-regression, i.e., it specifies how smooth the $\beta$ process is across neighbors.  
The conditional variance is a function of the number of neighbors, and intuitively these quantities are inversely proportional.
Though the $\beta$ process is not actually stationary, $\theta_\beta$ controls the \emph{conditional} precision: smaller $\theta_\beta$ leads to a larger variance.

%The $\bd\beta_g$ and $\bd\gamma_g$ processes are Markov and in the limit of infinitely dense node points would inheret the smoothness properties of a Gaussian Markov process, which means a very ``bouncy'' process, with a derivative that does not exist, similar to an Ohrenstein Uhlenbeck (OU) process.  However, the resulting process for $y_{\bd s}$ will inheret the continuity properties of an integrated OU process which has a continuous derivative and is much more representive of the actual $y_{\bd s}$ data.  \redud{FORMALIZE?? ELABORATE??}

\begin{comment}
From this specification it can be seen that the conditional mean and correlation of can be found using the following fact

\begin{theorem*}[Rue and Held 2005, Thm 2.3]
If $\bd x$ is a GMRF with mean vector $\bd\mu$ and precision matrix $\bd Q$, then
\begin{align*}
   E(x_i | \bd x_{-i})   &=   \mu_i - \frac{1}{Q_{ii}} \sum_{j : j \sim i} Q_{ij}(x_j - \mu_j)   \\
   Corr(x_i, x_j | \bd x_{-ij})   &=   \frac{ -Q_{ij} }{ \sqrt{Q_{ii}Q_{jj}} }
\end{align*}
\end{theorem*}

We say that $\beta_i$ and $\beta|j$ are \emph{in-grain neighbors} if $i \sim j$
Hence the set $\{j : j \sim i \}$ can be partitioned 
\end{comment}

\subsection{Remaining Priors and Hyperpriors}   \label{sec:OtherPriors}

Again, in order to get a Student-$t$ likelihood under the hierarchical specification, we need the latent variables to be distributed $iid$ inverse-gamma;  we also let the degrees of freedom vary:
\begin{align*}
   \omega_m  &\stackrel{iid}{\sim}  InvGam \left( \frac{df}{2}, \frac{df}{2} \right)  \quad \quad  (m = 1,\ldots,M)   \\
%   &=   \left( \frac{ (df/2)^{\frac{df}{2}} }{ \Gamma(df/2) } \right)^N  \left( \prod_{n=1}^N \omega_n \right)^{-\frac{df}{2} - 1} \exp\{ -\frac{df}{2} \sum_{n=1}^N \frac{1}{\omega_n} \}
   [df]   &\propto   \frac{1}{df^2}
\end{align*}
The other priors and hyperpriors are as follows
\begin{align*}
   \mu_g   &\stackrel{iid}{\sim}   N(\mu, \tau^2)   \quad \quad   \mu \sim N(\overline{y}, 100^2)   \quad   \tau^2 \sim InvGam(0.001, 0.001)   \\
   \sigma^2   &\sim   InvGam(0.001,0.001)   \\
   \phi_\beta   &\sim   logN \left( \ln 0.6 , \ 2(\ln 0.8 - \ln 0.6) \right)   \\
   \phi_\gamma   &\sim   logN \left( \ln 0.8 , \ 2(\ln 1.0 - \ln 0.8) \right)   \\
   \nu_\beta, \nu_\gamma   &\stackrel{iid}{\sim}   N(0, 8^2)   \hspace*{0.87in}
   \theta_\beta, \theta_\gamma   \stackrel{iid}{\sim}   Gam(0.001, \text{rate = } 0.001)   \\
   \kappa_\beta, \kappa_\gamma   &\stackrel{iid}{\sim}   Beta \left( 32/5, 8/5 \right)   \quad \quad
   \rho_\beta, \rho_\gamma   \stackrel{iid}{\sim}   Unif \left( -0.4 , 1 \right)
\end{align*}
The parameters of the log-normal distributions were chosen so that the prior median and mean of $\phi_\beta$ would be 0.6 and 0.8, and similarly, such that these quantities for $\phi_\gamma$ would be 0.8 and 1.0. 
The parameters of the beta distribution were chosen so that the prior mean and mode of the variables would be 0.8 and 0.9, implying somewhat smooth latent processes.
Using (\ref{eq:RhoBounds}), $\rho_\beta$ and $\rho_\gamma$ need to be greater than -0.75 and -0.5 (respectively), but using these as lower bounds in the uniform priors led to poor conditioning of the precision matrices, especially for moderately large $\kappa_\beta$ or $\kappa_\gamma$.  
%The value of 0.001 occurring in the various hyperpriors was arbitrary, but the abundance of data suggested this choice would be of little consequence.
Relatively diffuse priors were used for the remaining (hyper)parameters.   %%% Storlie

\section{Computational Details}   \label{sec:Computation}

We again point out that the impressive computational efficiency afforded by the use of \verb1R-INLA1 for GMRF models was not available to us for a few reasons.
First, the fact that $\phi_\beta$ and $\phi_\gamma$ are unknown means that the the design matrices $\bd X_b$ and $\bd X_c$ are not fixed (which is not currently supported by \verb1R-INLA1).
Second, the complexity of the prior precision for $\bd\beta$ and $\bd\gamma$ in (\ref{eq:PriorPrecision}) cannot easily be accommodated.
Last, the INLA approximation accuracy is expected to suffer due to the number of hyperparameters associated with the both of these random fields.
Hence, we use MCMC sampling to explore the posterior distribution of the full set of parameters.
In the next two subsections we give the updates within one iteration of the MCMC for the model defined by the likelihood (\ref{eq:Likelihood}) and the priors in Section \ref{sec:Priors}.  
We use a Metropolis-within-Gibbs approach whereby groups of parameters are updated using their full conditional distributions; a Metropolis step is used for those sets of parameters whose full conditional is not a known distribution.
Specific details about implementation are mentioned last.

%was accomplished through Markov chain Monte Carlo sampling of the posterior distribution of the complete set of parameters
%$$  \phi_\beta,  \nu_\beta,  \theta_\beta,  \rho_\beta,  \kappa_\beta, \quad
%   \phi_\gamma, \nu_\gamma, \theta_\gamma, \rho_\gamma, \kappa_\gamma, \quad
%   \bd\mu, \bd\beta, \bd\gamma  $$
%where $\bd\mu=(\mu_1,\ldots,\mu_G)$ is the vector of grain means.  

\subsection{Update of the Latent Fields and Hyperparameters}

As pointed out in \cite{Lee2002}, ``Finding an efficient method for updating the [Markov random] field turns out to be an interesting problem."
\cite{KnorrHeld2002} and \cite{Rue2005} Sec. 4.1.2 address this problem through the use of block updates; the authors report mixing within the MCMC that is superior to that produced by hybrid Gibbs updates, and for no additional computational cost.
Employing such a strategy in our case would entail the joint update of each random field and its hyperparameters, i.e. in two blocks:
first $(\bd\beta, \theta_\beta, \kappa_\beta, \rho_\beta)$ and then $(\bd\gamma, \theta_\gamma, \kappa_\gamma, \rho_\gamma)$.
The reasoning will be detailed after introducing some more notation; the discussion focuses on the $\beta$ field but applies directly to the $\gamma$ field as well.

Suppose that all hyperparameters of the unobserved random field $\bd\beta$ are collected in the vector $\bd\alpha_\beta$;
also let $[ \bd\beta^* | \bd\alpha_\beta^*]$ denote the density of the full conditional posterior $[ \bd\beta | \bd\alpha_\beta , \ldots , \bd y ]$ evaluated at $\bd\beta^*$ given $\bd\alpha_\beta^*$ and all other quantities.  
Within a Metropolis step, the jumping rule defined by
\begin{align}
  q_{\text{full}}\left( \bd\beta^*, \bd\alpha_\beta^*  |  \bd\beta^{(t-1)}, \bd\alpha_\beta^{(t-1)} \right)   \defi  q\left( \bd\alpha_\beta^*  |  \bd\alpha_\beta^{(t-1)} \right) \cdot [ \bd\beta^* | \bd\alpha_\beta^* ]   \label{eq:BlockUpdate}
\end{align}
has the same acceptance probability as the $\bd\alpha_\beta$ update alone because the density values associated with $\bd\beta$'s full conditional cancel in the acceptance ratio.

The preceding algorithm assumes that the entire field can be sampled at once, which is not the case for the current analysis. 
This is because the full conditional of $\bd\beta$ is normal with mean and covariance that depend on the inverse of $\bd X_b^\top \bd W \bd X_b + \bd Q_\beta$, which is a large dense matrix \citep[Equation 13]{Geweke1993}. 
It is also worth noting that even the formation of the full cross-product $\bd X_b^\top \bd W \bd X_b$ is not feasible because the design matrices are relatively dense.
We thus propose an extension of the block update strategy of \cite{Rue2005}, pg. 143.

Instead of proposing a candidate $\bd\beta^*$ from the density $[ \bd\beta | \bd\alpha_\beta , \ldots , \bd y ]$ (i.e. the entire vector all at once), our modification uses $S$ subblock proposals with a joint acceptance of the whole collection; the resulting Metropolis step then has jumping rule:
\begin{align}
   q_{\text{full}}\left( \bd\beta^*, \bd\alpha_\beta^*  |  \bd\beta^{(t-1)}, \bd\alpha_\beta^{(t-1)} \right)   &\defi  
   q\left( \bd\alpha_\beta^*  |  \bd\alpha_\beta^{(t-1)} \right)   \label{eq:SubblockUpdate}   \\
   &   \hspace*{0.5in}   \cdot [ \bd\beta_1^* | \bd\beta_2^{(t-1)}, \bd\beta_3^{(t-1)}, \ldots, \bd\beta_S^{(t-1)}, \bd\alpha_\beta^* ]   \nonumber   \\
   &   \hspace*{0.5in}   \cdot [ \bd\beta_2^* | \bd\beta_1^*, \quad \ \bd\beta_3^{(t-1)}, \ldots, \bd\beta_S^{(t-1)}, \bd\alpha_\beta^* ]   \nonumber   \\
   &   \quad \cdots \quad \cdot [ \bd\beta_S^* | \bd\beta_1^*, \quad \ \bd\beta_2^*, \quad \ \ldots, \bd\beta_{S-1}^*, \ \bd\alpha_\beta^* ]   \nonumber
\end{align}
The full conditional densities above are exactly those associated with sequential Gibbs updates for the corresponding blocks (the dependence upon $\bd y$ and the remaining parameters was again suppressed).
It can easily be seen that under this proposal rule, the conditional posterior terms will not cancel in the acceptance ratio meaning that $2S$ density evaluations are necessary for a complete update of $\bd\beta$: $S$ evaluations as prescribed by (\ref{eq:SubblockUpdate}), and $S$ more from the detailed balance computation, i.e., swapping the roles of ``$*$" and ``$(t-1)$".  
We have observed that this added computational cost is justified by superior exploration of the posterior.  
%\redud{say more somewhere???}
(Note: what we call ``blocks" Rue and Held call ``subblocks", implying that our ``subblocks" would be something like ``sub-subblocks" in their terminology.)
Let $\bd\beta_s$ and $\bd\beta_{\bar s}$ denote, respectively, the $s^{th}$ subblock and it's complement in $\bd\beta$.
%Typically $\bd\beta_k$ has hundreds of elements whereas $\bd\beta_{\bar k}$ has thousands.
Let other vectors/matrices be defined in a similar manner:
for example, 
$\bd X_s$ is the submatrix of $\bd X_b$ with columns corresponding to $\bd\beta_s$; 
$\bd Q_{s, \bar s}$ is the submatrix of the prior precision $\bd Q_\beta$ obtained by keeping rows and removing columns corresponding to the $s^{th}$ block of $\bd\beta$.

To compute the jumping rule in (\ref{eq:SubblockUpdate}), a closed form for $[ \bd\beta_s | \bd\beta_{\bar s}, \bd\alpha_\beta , \ldots , \bd y ]$ is necessary.  
Using properties of multivariate normal distributions \citep[Thm. 2.5]{Rue2005}  and some tedious but standard Bayesian calculations, it follows that
\begin{align}
   \bd\beta_s | \bd\beta_{\bar s}, \text{rest}   & \sim   N \left( \bd\nu_{s | \cdot}, \bd Q_{s | \cdot}^{-1} \right)   \\
   \bd\nu_{s | \cdot}   &\defi   \bd Q_{s | \cdot}^{-1}   \left( \bd X_s^\top \bd r_{\bar s}  +  \nu_\beta \bd Q_{s,s} \bd 1 - \bd Q_{s,\bar s} (\bd\beta_{\bar s} - \nu_\beta \bd 1) \right)   \nonumber   \\
   \bd Q_{s | \cdot}   &\defi   \bd X_s^\top \bd W \bd X_s + \bd Q_{s,s}   \nonumber   \\
   \bd r_{\bar s}   &\defi   \bd y  -  \bd\mu  -  \bd X_{\bar s} \bd\beta_{\bar s}  -  \bd X_c \bd\gamma   \nonumber
\end{align}
A few remarks are in order.
Despite $\bd Q_\beta$ being sparse, $\bd Q_{s|\cdot}$ is dense because of the cross-product term.
However, the inverse of this posterior precision is manageable because $\bd\beta_s$ is specified to give $\bd X_s$ a reasonable number of columns.
This implies that $\bd\beta_{\bar s}$ is quite large, but the sparsity of the prior precision makes $\bd Q_{s,\bar s} (\bd\beta_{\bar s} - \nu_\beta \bd 1)$ easy to compute.
The partial residual $\bd r_{\bar s}$ depends on a potentially unwieldy $\bd X_{\bar s} \bd\beta_{\bar s}$, but this can be efficiently managed by updating the full residual 
\begin{align}
   \bd r   =   \bd y - \widetilde{\bd\mu}   \ = \   \bd y - \bd\mu - \bd X_b \bd\beta - \bd X_c \bd\gamma   \label{eq:FullResidual}
\end{align}
for each of the $S$ subblocks.
To update the $s^{th}$ subblock of $\bd\beta$, set $\bd r_s = \bd r + \bd X_s \bd\beta_s$, draw the new subblock $\bd\beta_s^*$, and then update 
%$\bd r \leftarrow \bd r + \bd X_s(\bd\beta_s - \bd\beta_s^*)$ 
$\bd r \leftarrow \bd r_s - \bd X_s \bd\beta_s^*$ 
and $\bd\beta_s \leftarrow \bd\beta_s^*$.
Note that to evaluate the density and obtain a new draw from the full conditional only one Cholesky decomposition of $\bd Q_{s | \cdot}$ is needed.

The other term necessary to compute the jumping rule in (\ref{eq:SubblockUpdate}) is $q( \bd\alpha_\beta^*  |  \bd\alpha_\beta^{(t-1)} )$.
For this function we use a multivariate normal density (centered on the previous value) in the transformed space
\begin{align*}
   \bd\alpha_\beta   \defi   \left( \ln(\phi_\beta), \   \ln\left(\frac{\theta_\beta}{\kappa_\beta}\right), \   \Phi^{-1}(\kappa_\beta), \   \Phi^{-1}\left( \frac{\rho_\beta + 0.4}{1 + 0.4} \right) \right) 
\end{align*} 
Above, $\Phi^{-1}(\cdot)$ is the quantile function of a standard normal variate and the ``0.4" and ``1" terms come from uniform prior on $\rho_\beta$; by design, all transformed variables have unbounded support.
The resulting $q(\cdot | \cdot)$ is symmetric and will thus cancel in the joint Metropolis acceptance ratio, but the prior densities must be adjusted according to a change-of-variables.
%Transforming the hyperparameters and the inclusion of $\phi_\beta$ into the joint update were found to make a substantive positive difference on the performance of the MCMC.
Transforming the hyperparameters and the inclusion of $\phi_\beta$ into the joint update were essential adjustments to allow for adequate mixing of the MCMC algorithm.   %%% Storlie

\subsection{Remaining Updates}

The last parameter of the $\beta$ process which is not updated jointly with $\bd\beta$ has full conditional distribution
\begin{align*}
   \nu_\beta | \text{rest}   &\sim   N \left( \frac{\bd 1^\top \bd Q_\beta \bd\beta + 0/(8^2)}{\bd 1^\top \bd Q_\beta \bd 1 + 1/(8^2)} \ , \ \frac{1}{\bd 1^\top \bd Q_\beta \bd 1 + 1/(8^2)} \right)  
\end{align*}
0 and $8^2$ being the prior mean and variance.
In what follows, $\bd r$ again denotes the full residual (\ref{eq:FullResidual}), with $mth$ entry $r_m$.
The parameters associated with the grain means have full conditionals

\begin{align*}
   \mu_g | \text{rest}   &\stackrel{iid}{\sim}   N \left( \text{mean} = \left( \frac{1}{\sigma^2} \sum_{g(m)=g} \frac{r_m}{\omega_m} + \frac{\mu}{\tau^2} \right) \cdot \text{var} \ ,   \right.   \\
   &   \hspace*{0.62in} \left.   \text{var} = \left( \frac{1}{\sigma^2} \sum_{g(m)=g} \frac{1}{\omega_m} + \frac{1}{\tau^2}\right)^{-1} \right)   \\
   \mu | \text{rest}   &\sim   N \left( \left( \frac{\sum_{g=1}^G \mu_g}{\tau^2} + \frac{\overline y}{100^2} \right) \cdot \text{var} \ , \ \text{var}= \left(\frac{G}{\tau^2} + \frac{1}{100^2} \right)^{-1} \right)   \\
   \tau^2 | \text{rest}   &\sim   InvGam \left( \frac{G}{2} + 0.001 \ , \ \frac{1}{2} \sum_{g=1}^G (\mu_g - \mu)^2 + 0.001 \right)   
\end{align*}
The parameters of the error distribution of $\bd y$ have full conditionals
\begin{align*}
   \sigma^2 | \text{rest}   &\sim   InvGam \left( \frac{M}{2} + 0.001 \ , \ \frac{1}{2} \sum_{m=1}^M \frac{r_m^2}{\omega_m} + 0.001 \right)   \\
   \bd\omega | \text{rest}   &\stackrel{iid}{\sim}   InvGam \left( \frac{1}{2} + \frac{df}{2} \ , \ \frac{r_m^2}{2\sigma^2} + \frac{df}{2} \right)   \\
   [df | \text{rest}]   &\propto   \Gamma \left( \frac{df}{2} \right)^{-M} \left( \frac{df}{2} \right)^{M df/2 - 2} \left( \prod_{m=1}^M \frac{1}{\omega_m} \right)^{df/2} \exp\left( - \frac{df}{2} \sum_{m=1}^M \frac{1}{\omega_m} \right)
\end{align*}
The $df$ parameter can easily be updated using a univariate normal density as a jumping rule.
The updates provided in this section are computationally trivial and represent a tiny percentage of the total time associated with one iteration of the MCMC.

\begin{figure}[h!t]   \begin{center}
\includegraphics[width=4.0in]{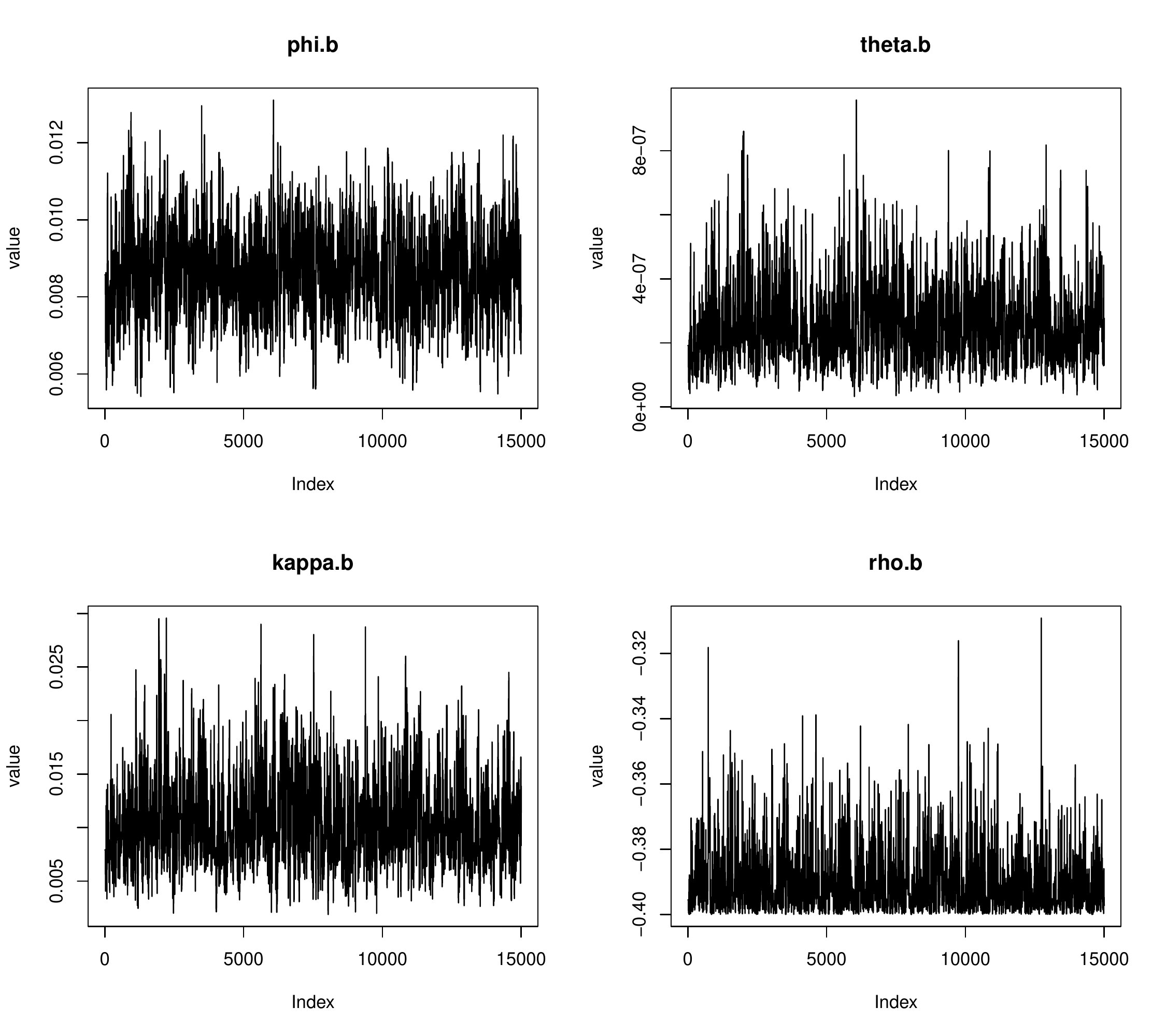} 
\caption{ Trace plots (after discarding the burn-in iterations and then back-transforming) for one parameter and three hyperparameters associated with the $\beta$ process using grains as subblocks. }
\label{fig:TracePlots}
\end{center}   \end{figure}

\subsection{Remaining Details of Implementation}

The choice of subblocks affects the performance of the MCMC routine.
%Heuristically, one should use subblocks which are expected to be correlated and as large as possible to gain the sampling efficiency of the joint update strategy (\ref{eq:BlockUpdate}).
Heuristically, one should use subblocks which are expected to be correlated within, minimally correlated between, and as large as possible to gain the sampling efficiency of the joint update strategy (\ref{eq:BlockUpdate}).   %%% Storlie
On the other hand, using smaller parameter patches can be faster computationally.
For each of the $\bd\beta$ and $\bd\gamma$ vectors we were able to use one subblock per grain; this led to a maximum $\bd X_g$ dimension of $17946 \times 2784$ for $g=11$.
We found that subblocks of roughly 200 coefficients was faster but produced chains which wandered a bit more. 
%We found that subblocks of roughly 200 (leading to \qqq total subblocks) was faster but produced chains which wandered a bit more. 
%Further study of this trade-off\qqq is needed.
%We hypothesize that the use of alternating subblocks?? the mixing will be better  \qqq

The choice of hardware and software also affect computational performance.
In this study we used a Mac Pro desktop with a 3.5 GHz, 6-Core Intel Xeon E5 processor and 32GB of memory.
To fully harness the multicore capabilities, \verb1R1 was compiled against OpenBLAS 0.2.18 to provide default parallelization.
This was necessary for two reasons.
First, the parameters $\phi_\beta$ and $\phi_\gamma$ were not fixed which implies that each iteration involves multiplying and exponentiating $ >1.20 \ttplus 8$ terms for the $\bd X_b$ matrix and $>2.57 \ttplus 7$ for the $\bd X_c$ matrix-- even automated parallel matrix algebra can buckle under such a burden.
Second, to use one subblock per grain, a matrix cross-product and Cholesky decomposition involving the large $\bd X_{11}$ is necessary.
Parallelized matrix linear algebra is indispensable in such a scenario.
Sparse matrices were handled using the \verb1Matrix1 package within \verb1R1 3.2; it was found that this particular software outperformed the functionality provided by both \verb1spam1 and \verb1SparseM1.
Even with the considerable gains derived from the careful implementation within \verb1R1, a full round of updates still takes about 16 seconds.  
Using smaller blocks, one iteration could be sped up to $\sim$6 seconds, but we opted for bigger blocks to aid sampling efficiency.

The parameters requiring a multivariate normal Metropolis jumping rule, i.e. $\bd\alpha_\beta$,  $\bd\alpha_\gamma$ and $df$, had their proposal covariances tuned during the burn-in period according to the method of \cite{Haario1999}.
% \cite{Haario2001} \cite{Roberts2009}
After a block of 500 iterations, the covariance was adjusted by the appropriate multiplier necessary to get an acceptance rate of $\approx 0.234$, as suggested by \cite{Gelman1996}; this was done for 20 such blocks.
An additional 5000 samples from fixed proposal covariances were used for the burn-in period, implying a total of 15,000 not used in the final analysis.
The MCMC routine was run for 15,000 more iterations, and every fifth sample was recorded.

\section{Results}   \label{sec:Results}

\begin{figure}[h!t]   \begin{center}
\includegraphics[width=3.2in]{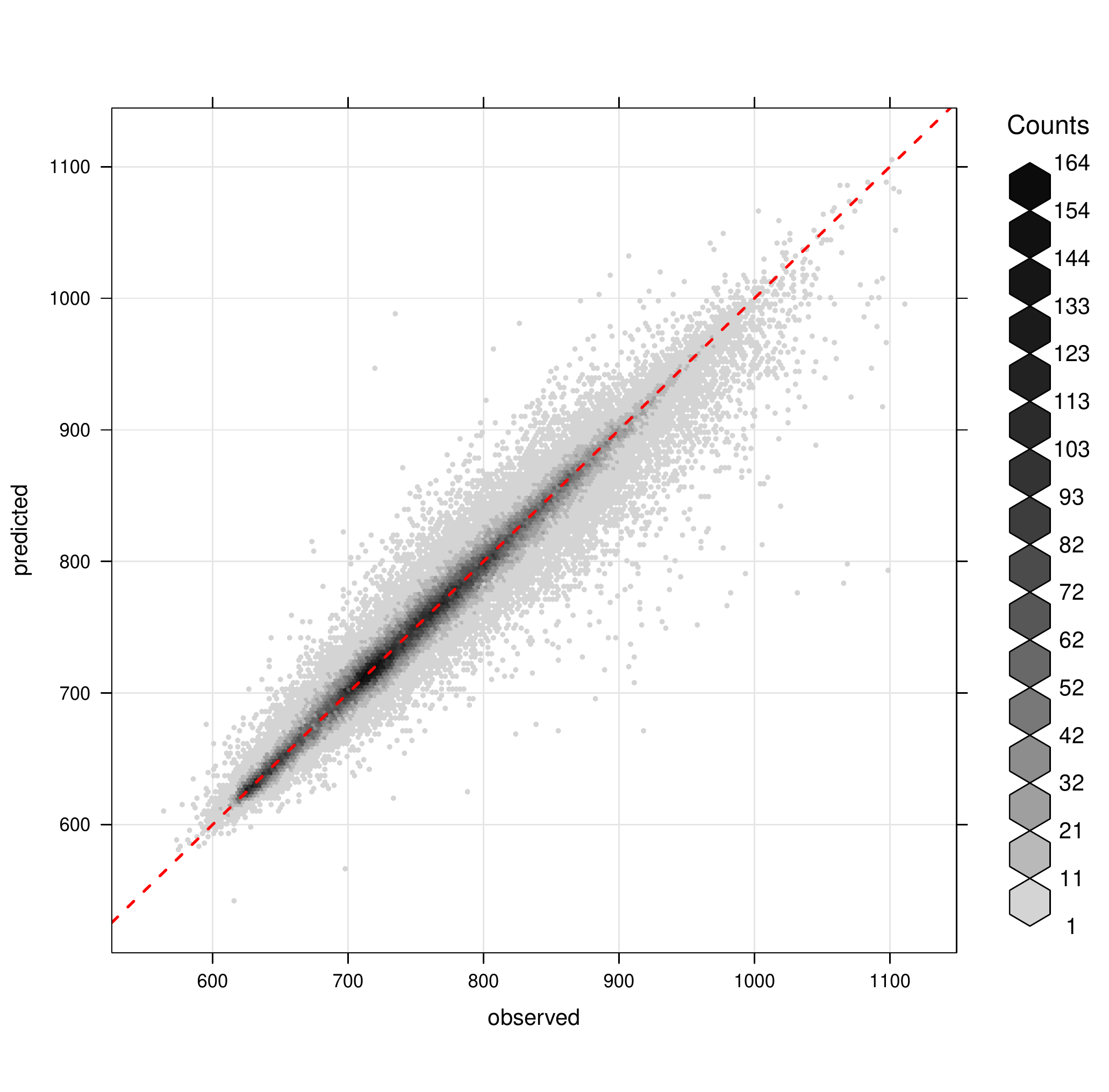} 
\caption{ Hexbin plot of predicted versus observed for the last iteration of the MCMC; the red dashed one-to-one line is displayed for visual aid. }
\label{fig:ObsVsPred}
\end{center}   \end{figure}

\begin{figure}[h!t]   \begin{center}
\includegraphics[width=5.0in]{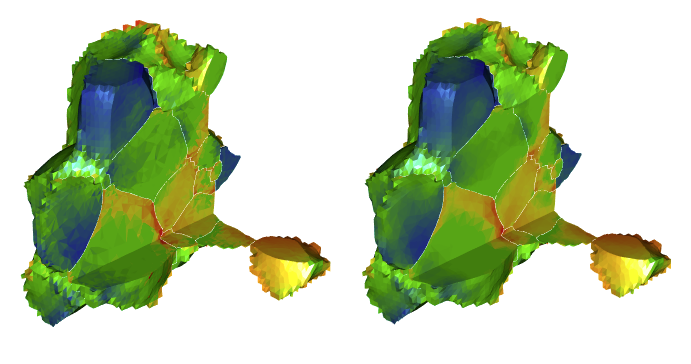} 
%\hspace*{-1.0in}
%\includegraphics[width=5.0in]{18_grain-cutout.pdf} 
%\hspace*{-2.7in}
%\includegraphics[width=5.0in]{18_grain_pred-cutout.pdf} 
%%\hspace*{-2.0in}
%\includegraphics[width=4.1in]{18_grain_pred.pdf} 
\caption{ Cutouts of the 18 grain domain with observed vonMises stress (left) and predicted values (right) for the last iteration of the MCMC. }
\label{fig:ObsVsPredGrains}
\end{center}   \end{figure}

\begin{figure}[h!t]   \begin{center}
\includegraphics[width=5.0in]{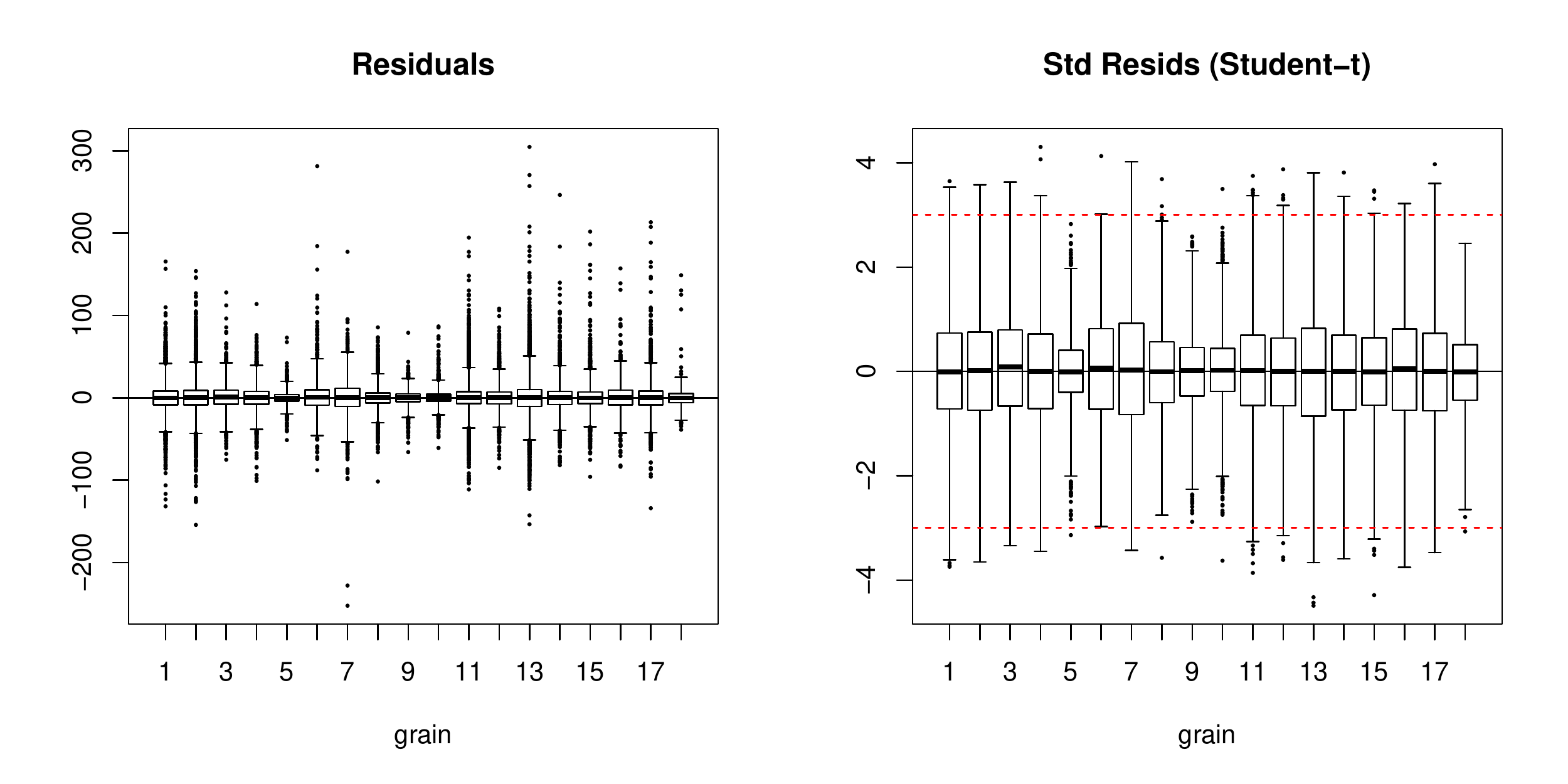} 
\caption{ Raw and standardized residuals for the last iteration of the MCMC. }
\label{fig:Residuals}
\end{center}   \end{figure}

\begin{figure}[h!t]   \begin{center}
\includegraphics[width=5.0in]{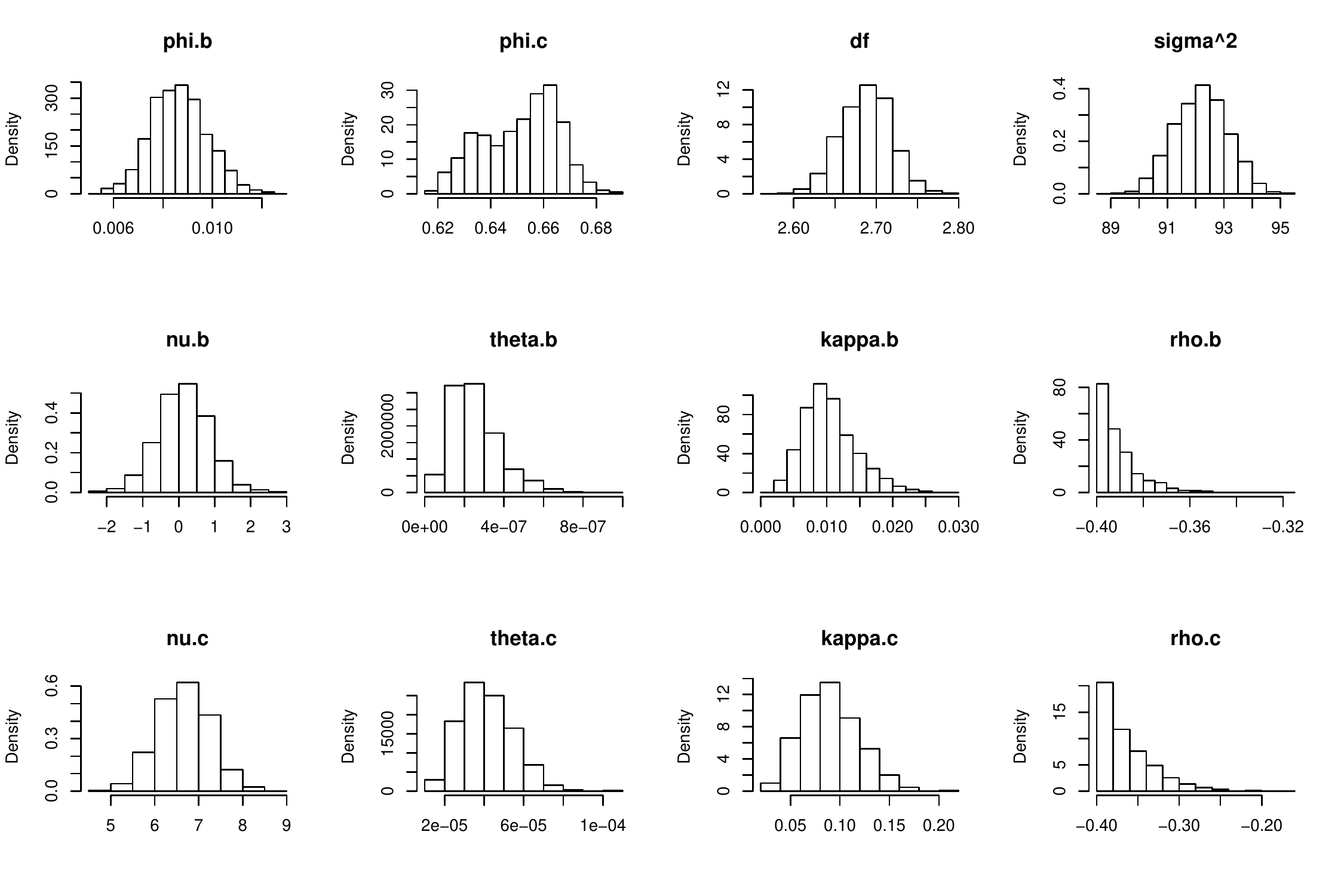} 
\caption{ One dimensional marginal posterior distributions for various parameters. }
\label{fig:MarginalPosteriors}
\end{center}   \end{figure}

Aside from the 94,274 auxiliary variables of $\bd\omega$ necessary for the Student-$t$ likelihood, there are a total of $dim(\bd\beta) + dim(\bd\gamma) + G + 14 = 17,948$ parameters/hyperparameters.
Of these, the convergence of the random fields $\bd\beta$ and $\bd\gamma$ is somewhat hard to assess, but we did not observe anything conspicuously egregious in the trace plots of randomly selected entries of the vectors.
Trace plots for the hyperparameters of the $\beta$ process are shown in Figure \ref{fig:TracePlots}.
The (qualitative) adequacy of the chains' character is a direct result of our sampling routine.

Every fifth iteration of the MCMC we monitored some quantities to assess the model's goodness-of-fit.
For example, adjusted coefficients of determination $R^2_{adj}$ were calculated; 
after the burn-in period, $R^2_{adj}$ relative to a constant mean model was above 0.98 and the 
$R^2_{adj}$ relative to a grain-specific mean model was above 0.95.

For one such iteration (the last of the MCMC), additional diagnostic plots are given in Figures \ref{fig:ObsVsPred}, \ref{fig:ObsVsPredGrains}, and \ref{fig:Residuals}.
(The reason for using only one iteration, as opposed to say the mean field $\widehat{\bd\mu} + \widehat{\bd X_b \bd\beta} + \widehat{\bd X_c \bd\gamma}$, is the presence of $\bd\omega$: the mean field will not correspond to the mean of the auxiliary variables $\widehat{\bd\omega}$, and this will invalidate residual-based diagnostics.)
% so why not use the MAP?? ... because I didn't store the total posterior values
The plots comparing observed and predicted values corroborate the high $R^2$ values to suggest that the model fits the data well, even adjusting for the huge number of parameters.
We did not use any cross-validation based diagnostics for two reasons.
First, the additional computations for a $k$-fold cross-validation were deemed undesirable, given the already taxing MCMC.  
Second, and more importantly, using a hold-out subset of the data for validation purposes will almost certainly not capture the relevant generalization error: that which is associated with data on new/different grain geometries.
This type of model assessment can only be derived from verification with further simulated datasets.
% \qqq Another measure of fit on *new* dataset...  use phi's and hypers, use gibbs to get beta,gamma,omega and look at adjusted R2

Figure \ref{fig:Residuals} shows the residuals for the last iteration of the MCMC.
The raw residuals indicate some extreme absolute errors relative to the range of the data, but these are indeed a small minority.
They also display inappropriate tail behavior implying that any constant-variance normal error model will not be suitable.
The left panel stands in contrast to the right panel of Figure \ref{fig:Residuals} which favors the use of a Student-$t$ likelihood (the $m$th residual is standardized by $\sqrt{\omega_m}\sigma$).
We note however that a further examination of the residual field reveals some potential spatial correlation, especially near grain boundaries; this will be an avenue for future investigation and improvement.

Marginal posterior distributions for some of the parameters are displayed in Figure \ref{fig:MarginalPosteriors}.  
A few noteworthy features are present.
First, the small values of $df$ are further evidence for a heavy-tailed error distribution.
Next, the quantity $1/\phi_\beta$ is related to a correlation length, and in this light the reciprocal of the posterior mode is very large: the effect of the $\beta$ process ``reaches across" the entirety of each grain.
The $\beta$ process mean $\nu_\beta$ is centered on zero, meaning that a second-order boundary can either elevate or lower expected vonMises stress; $\nu_\gamma > 0$ suggests that third-order boundaries are more often tied to the elevation of stress.
Both $\kappa$ parameters are small implying that the processes are actually quite rough, and plots of $\bd\beta$ and $\bd\gamma$ (not presented here) were visually indistinguishable from noise. 
Both $\rho$ parameters are negative implying an inhibitory effect within collection of $\{\bd\beta_g\}$ and $\{\bd\gamma_g\}$.
This means that the values of two processes on the same grain boundary are negatively correlated (in a conditional sense); they are not in fact independent.
The posteriors of both $\rho$ parameters push up against the lower bounds of the prior which is undesirable as well as unexpected-- it provides some evidence of model inadequacy despite the favorable diagnostics presented earlier.  

\section{Conclusions}

This paper represents a first step towards understanding vonMises stress fields and hence damage initiation within polycrystalline tantalum, a process of great interest to scientists at Los Alamos National Laboratory.
Because the constitutive mechanical equations dictating the material response did not satisfactorily describe the visible spatial variability to the materials scientists, we proposed an empirical statistical model for the complicated, high-dimensional and rich simulated tantalum dataset. 
The data's size and tetrahedrally meshed geometry (hence a pre-existing neighborhood structure), strongly motivated the use of Gaussian Markov random fields.
However, within our unique model, the variability throughout the entire 3D volume was dictated by latent and interacting GMRFs defined on lower dimensional 2D and 1D grain boundaries.
As such, the scientists' intuition of grain boundary importance was built directly into the model.
We also allowed for a heavy-tailed error distribution so that outliers did not have undue influence.

Our novel GMRF model required a careful Bayesian implementation and we proposed the use of a modified block updating scheme for the latent fields.
Sparse matrix functionality and parallel computing methods were vital to the performance of the MCMC routine.

After fitting the model we encountered some surprising results.
We had expected the latent processes at the second- and third-order boundaries (the $\bd\beta$ and $\bd\gamma$ fields) to be smooth (large $\kappa_\cdot$ hyperparameters), and their influence to have a moderate to quick decay away from the boundaries (indicated by $\phi_\cdot$ hyperparameters).
However, the data unequivocally encouraged extremely rough latent fields that, through small decay parameters, averaged out to a smooth 3D mean field that was visually quite similar to the observed.
Also, despite good visual agreement between observed and predicted, the $\rho_\cdot$ parameters had modes at the bounds required to guarantee precision matrix invertibility. 
This means that, loosely speaking, the model had to ``stretch" to accommodate the type of variation present in the data.
On the other hand, the roughness of the latent fields implies a large \emph{effective} number of parameters (in $\bd\beta$ and $\bd\gamma$) which is a sign of overfitting.
This means that more thought needs to be given to how the latent fields can be smoothed (``regularized") in a principled way beyond our current hyperparameter specification.

Past exploring the issues mentioned in the previous paragraph, there is considerable room for improving this (or any) statistical model of stress within simulated shock-loaded tantalum.
For instance, future work will have to accommodate substantially larger datasets (on the order of many millions of spatial locations), tensor-valued output, and grain orientation as a covariate.
Another major issue to consider is the dependence of hyperparameters upon the given tetrahedral meshing.
Mesh-invariant model specification is a difficult problem from the GMRF perspective (see e.g., \cite{Besag2005} and \cite{Lindgren2011}), but one whose solution would allow for better generalizability and predictions upon new grain geometries with arbitrary mesh structure.
A possibility is to ignore the mesh geometry and instead favor notions of distance and correlation length, but one is then immediately back to the problem of specifying a non-stationary model for big spatial data.
The difficulties within both mesh-based or distance-based approaches makes this an important and challenging area of research and application.
%
%We will also pursue models which do not use integrals and which feature hyperparameters with more graceful interpretations.

\bibliographystyle{asa}

\bibliography{0-Biblio}

\end{document}